\documentclass[12pt,letterpaper]{article}

\usepackage{verbatim}
\usepackage{float}
\usepackage{booktabs}
\usepackage{amsmath}
\usepackage{amssymb}
\usepackage{graphicx}
\usepackage{setspace}
\usepackage{natbib}
\usepackage{algorithm}
\usepackage{algorithmic}
\usepackage{xcolor, colortbl,epstopdf}

\definecolor{lightgray}{gray}{0.9}

\bibpunct{(}{)}{;}{a}{,}{,}

\newcommand{\ba}{\mathbf{a}}
\newcommand{\bA}{\mathbf{A}}

\newcommand{\bB}{\mathbf{B}}

\newcommand{\bD}{\mathbf{D}}

\newcommand{\bG}{\mathbf{G}}

\newcommand{\bh}{\mathbf{h}}

\newcommand{\bL}{\mathbf{L}}

\newcommand{\bu}{\mathbf{u}}

\newcommand{\by}{\mathbf{y}}

\newcommand{\bV}{\mathbf{V}}

\newcommand{\vect}[1]{\boldsymbol #1}

\newcommand{\valpha}{\vect{\alpha}}

\newcommand{\vLambda}{\vect{\Lambda}}

\newcommand{\vepsilon}{\vect{\epsilon}}

\newcommand{\vSigma}{\vect{\Sigma}}

\newcommand{\vxi}{\vect{\xi}}

\newcommand{\e}{\text{e}}
\newcommand{\gvn}{\,|\,}
\renewcommand{\epsilon}{\varepsilon}

\renewcommand{\leq}{\leqslant}

\newcommand{\distn}[1]{\mathcal{#1}}

\begin{document}

\title{BVARs and Stochastic Volatility}
\author{Joshua C. C. Chan\thanks{This review is prepared for the \textit{Handbook of Macroeconomic Forecasting} edited by Mike Clements and Ana Galvao. I thank Todd Clark, Frank Wu,  Wei Zhang and the editors Mike Clements and Ana Galvao for their helpful comments and suggestions.} \\
Purdue University \\
\footnotesize{ORCID: 0000-0003-3632-128X}
}

\date{October 2023}

\maketitle

\begin{abstract}

\noindent Bayesian vector autoregressions (BVARs) are the workhorse in macroeconomic forecasting. Research in the last decade has established the importance of allowing time-varying volatility to capture both secular and cyclical variations in macroeconomic uncertainty. This recognition, together with the growing availability of large datasets, has propelled a surge in recent research in building stochastic volatility models suitable for large BVARs. Some of these new models are also equipped with additional features that are especially desirable for large systems, such as order invariance---i.e., estimates are not dependent on how the variables are ordered in the BVAR---and robustness against COVID-19 outliers. Estimation of these large, flexible models is made possible by the recently developed equation-by-equation approach that drastically reduces the computational cost of estimating large systems. Despite these recent advances, there remains much ongoing work, such as the development of parsimonious approaches for time-varying coefficients and other types of nonlinearities in large BVARs.


\smallskip

\noindent \textbf{Keywords}: large Bayesian VAR, heavy tail, heteroskedasticity, forecasting, order invariance, COVID-19

\end{abstract}

\thispagestyle{empty}

\newpage

\section{Introduction}

Bayesian vector autoregressions (BVARs) are the workhorse in empirical macroeconomics, especially for forecasting applications and structural analysis. Since the pathbreaking work by \citet{sims80} and \citet{DLS84} in the early 1980s, there have been many important methodological advances in the BVAR literature. The most notable are the following two developments that motivate a large and expanding body of research in the last decade or so.

First, there is a growing interest in exploiting richer information in macroeconomic analysis, following the seminal paper by \citet*{BGR10} that demonstrates the benefits of including a large number of variables in BVARs.\footnote{The earlier paper by \citet{LSZ96} develops various medium-sized structural VARs to study the effects of monetary policy, including one with 18 variables. \cite{koop13} compares the forecast performance of a range of shrinkage priors for BVARs using a dataset containing 168 variables, and finds that BVARs tend to forecast better than factor models.} Indeed, there are many situations in which it is necessary to consider many variables simultaneously. For example, central banks and policy institutions routinely monitor and forecast dozens of key macroeconomic variables \citep{Crumpetal21}. In applications involving regional time-series, where a region can be a country within an economic union or a state/province within a country, it is often essential to explicitly model the interactions among the regions \citep{CC09, KK16, KMM20}. For nowcasting applications, it is important to incorporate the flow of data releases in real time, which necessitates a data-rich approach \citep{GRS08, BGMR13}.

Many other applications also naturally call for the inclusion of a large number of time-series, such as those involving data in multiple frequencies \citep{SS15, MOS21}, disaggregated data \citep{GLMO14, ER17}, firm-level data \citep{DDLY18} and financial data \citep{CKM09}. Finally, the wide availability of large time-series datasets \citep{MN16,MN20,BLS22,BJKO23} further propels this development.

The second development is the increasing recognition of the need to allow for time-varying volatility in modeling most macroeconomic datasets. Influential papers by \citet{CS05}, \citet{Primiceri05} and \citet{SZ06} underscore the secular variations in volatility---e.g., many key macroeconomic variables were more volatile in the Great Inflation of the 1970s than the following Great Moderation. At the business cycle frequency, a large body of evidence has shown that macroeconomic volatility is strongly counter-cyclical \citep[see, e.g.,][]{Bloom14, JLN15}. Lastly, the unexpected drastic movements in many macroeconomic variables at the onset of the COVID-19 pandemic and the subsequent heightened volatility further underline the need to allow for time-varying volatility. 

Within the BVAR literature, papers such as \citet{clark11}, \citet*{DGG13}, \citet{CR15}, \citet{CP16} and \citet{CE18} highlight the empirical relevance of time-varying volatility for model-fit and forecasting in small BVARs. For large BVARs, \citet{CCM16} and \citet{chan23JE} present Bayesian model comparison results that show overwhelming data support in favor of stochastic volatility. \citet*{KK13} and \citet{CCM19} demonstrate that BVARs with stochastic volatility forecast better than their homoskedastic counterparts. \citet{CM23} provide a recent review on the wide range of applications of BVARs with stochastic volatility.

The convergence of these two developments drives a surge in recent research in building stochastic volatility models suitable for large BVARs and designing efficient methods for estimating these models. We first introduce in Section~\ref{s:models} three classes of stochastic volatility models---the common stochastic volatility, the Cholesky stochastic volatility and the factor stochastic volatility---that are especially suitable for large BVARs. We then dive into some recent research in developing BVARs with features that are particularly desirable for large systems in Section~\ref{s:OIOA}. These include order invariance---i.e., estimation results that are not dependent on how the variables are ordered in the BVAR---and robustness against COVID-19 outliers. Section~\ref{s:estimation} first reviews Bayesian shrinkage priors, particularly the family of Minnesota priors. It then outlines Markov chain Monte Carlo (MCMC) methods for estimating BVARs with various types of stochastic volatility, followed by a discussion on alternative Bayesian approaches and additional strategies to speed up computations. Section~\ref{s:nonlinear} further explores BVARs with time-varying parameters and other types of nonlinearities. Finally, Section~\ref{s:conclusions} highlights a few outstanding challenges and ongoing research directions.


\section{Stochastic Volatility Models for VARs} \label{s:models}

In this section we discuss a few commonly-used stochastic volatility models in empirical studies involving macroeconomic forecasting and structural analysis. Since there is an increasing demand to incorporate a large number of time-series, we pay special attention to the trade-off between model flexibility and tractability in larger systems. 

Let $\by_t=(y_{1,t},\ldots, y_{n,t})'$ denote the $n$-vector of endogenous variables for $t=1,\ldots, T$. Consider the following VAR with a generic time-varying error covariance matrix $\vSigma_t$:
\begin{equation} \label{eq:VAR} 
	\by_t = \ba + \bA_1 \by_{t-1} + \cdots + \bA_p \by_{t-p} + \vepsilon_t, \quad \vepsilon_t \sim\distn{N}(\mathbf{0}_n,\vSigma_t),
\end{equation}
where $\ba$ is an $n\times 1$ vector of intercepts, $\bA_1,\ldots, \bA_p$ are $n\times n$ coefficient matrices and $\mathbf{0}_n$ denotes an $n$-vector of zeros. To model volatility clustering---e.g., the empirical observation that large changes tend to be followed by large changes, and small changes followed by small changes---it is important to specify a persistent law of motion for $\vSigma_t$. While there are many ways to construct persistent processes, the key challenge here is that for any time $t$, $\vSigma_t$ needs to be a positive-definite covariance matrix. In what follows, we discuss various models for $\vSigma_t$ that satisfy these requirements.

\subsection{Common Stochastic Volatility}

We first consider the common stochastic volatility model introduced in \citet*{CCM16}, which is the very first stochastic volatility model designed for large BVARs. The model is motivated by the observation that the estimated time-varying volatilities of many US macroeconomic variables have broadly similar low-frequency movements: they tend to have high volatility in the 1970s, a drastic decrease in volatility starting in the early 1980s that mark the beginning of the Great Moderation, followed by a new surge in volatility at the onset of the Great Recession of 2007-2009. In particular, \citet*{CCM16} obtain the estimated volatilities of 14 major US macroeconomic time-series from univariate autoregressive models with stochastic volatility, and find that the first principal component explains about 66\% of the variation in the individual volatility time-series. 

Motivated by this common component in volatilities, \citet*{CCM16} propose the common stochastic volatility model in which the error covariance matrix is scaled by a common, time-varying factor that may be interpreted as the overall macroeconomic volatility:
\begin{equation}\label{eq:csv}
	\vSigma_t = \e^{h_t}\vSigma, 
\end{equation}
where $\vSigma$ is a time-invariant covariance matrix. The log-volatility $h_t$ in turn follows a stationary AR(1) process:
\begin{equation}\label{eq:h}
	h_t = \phi h_{t-1} + u_t^h, \quad u_t^h\sim\distn{N}(0,\sigma^2),
\end{equation}
for $t=2,\ldots, T$, where $|\phi|<1$ and the initial condition is specified as 
$h_{1}\sim\distn{N}(0,\sigma^2/(1-\phi^2))$. Note that for identification purposes, the unconditional mean of the AR(1) process is assumed to be zero.

For large systems, estimation of the generic VAR with stochastic volatility in \eqref{eq:VAR} is computationally intensive because of the large number of VAR coefficients $\bA = (\ba_0, \bA_1, \ldots, \bA_p)'$: it grows quadratically in the number of endogenous variables $n$. The key advantage of the common stochastic volatility model is that it leads to many useful analytical results that make estimation fast, with computational complexity of the order $\mathcal{O}(n^3)$---provided that the natural conjugate prior on $(\bA, \vSigma)$ is used \citep[see, e.g.,][and Chapter 2 of this volume by Hauzenberger, Huber and Koop]{KK10, karlsson13}. Consequently, estimation of BVARs with dozens endogenous variables can be done in minutes instead of hours for other multivariate stochastic volatility models. 

The main drawback of the common stochastic volatility model is that it appears to be restrictive along two dimensions. First, fast estimation with computational complexity $\mathcal{O}(n^3)$ is only possible when the natural conjugate prior is used, but the natural conjugate prior can be restrictive as it rules out cross-variable shrinkage \citep[see, e.g.][]{CCM15, chan22}. Second, the common stochastic volatility model assumes in particular that all variances and covariances are scaled by a single factor. Consequently, they are always proportional to each other. Nevertheless, as discussed above, there is empirical evidence that the error variances of many US macroeconomic variables share broadly similar low-frequency movements, and a common stochastic volatility is a parsimonious way to model this empirical feature.

The common stochastic volatility model in \eqref{eq:csv}--\eqref{eq:h} can be extended in various ways. In particular, \citet{chan20} considers a more general family of BVARs that can accommodate non-Gaussian, heteroskedastic and serially dependent errors. In addition, fast estimation of complexity 
$\mathcal{O}(n^3)$ can be attained for this wide family of BVARs. 

Recent empirical applications using this common stochastic volatility model and its extensions include \citet{muntaz16}, \citet{MT17}, \citet{GH18}, \citet{Poon18}, \citet{Louzis19}, \citet{LH19}, \citet{Hartwig21}, \citet{CPZ23} and \citet{HNZ23}.

\subsection{Cholesky Stochastic Volatility}

Instead of scaling the error covariance matrix by a common volatility factor, a more flexible approach for modeling time-varying volatilities and correlations is to incorporate multiple stochastic volatility processes. In particular, \citet{CS05} build a multivariate stochastic volatility model based on the modified Cholesky decomposition. More specifically, consider again the VAR in \eqref{eq:VAR}, but now the error covariance matrix is constructed via
\begin{equation} \label{eq:VAR-SV}
	\vSigma_t = \bB_0^{-1} \bD_t (\bB_0^{-1})',
\end{equation}
where $\bB_{0}$ is an $n \times n$ lower triangular matrix with ones on the diagonal and $\bD_t = \text{diag}(\e^{h_{1,t}}, \ldots, \e^{h_{n,t}})$. The law of motion for each element of $\bh_t = (h_{1,t}, \ldots, h_{n,t})'$ is specified as an independent autoregressive process:
\begin{equation}\label{eq:hit}
	h_{i,t} = \mu_i + \phi_i(h_{i,t-1}-\mu_i) + u_{i,t}^h, \quad u_{i,t}^h \sim \distn{N}(0, \sigma_{i}^2)
\end{equation}
for $t=2,\ldots, T$, where the initial condition is specified as $h_{i,1}\sim\distn{N}(\mu_i,\sigma_i^2/(1-\phi_i^2))$. Due to its construction using the modified Cholesky decomposition, this stochastic volatility model is sometimes called the Cholesky stochastic volatility. 

Comparing to the common stochastic volatility model in \eqref{eq:csv}, it is clear that the Cholesky stochastic volatility is more flexible---it contains $n$ stochastic volatility processes---and can accommodate more complex co-volatility patterns. In particular, the Cholesky stochastic volatility can in principle recover the common stochastic volatility specification if we reparameterize the model in \eqref{eq:VAR-SV} so that $\bB_0$ is the lower Cholesky factor of $\vSigma^{-1}$ and $\bD_t = \e^{h_t}\mathbf{I}_n$, where $\mathbf{I}_n$ is the $n$-dimensional identity matrix. 

This modeling flexibility, however, comes at a cost of more computationally intensive posterior simulations. In particular, the conventional approach of drawing all VAR coefficients jointly, which has $\mathcal{O}(n^6)$ computational complexity, becomes excessively time-consuming when $n$ is large. To tackle this computational problem, \citet*{CCM19} introduce a blocking scheme that makes it possible to estimate the model equation by equation that drastically reduces the computational complexity to $\mathcal{O}(n^4)$; see Section~\ref{s:estimation} for more details. Nevertheless, the estimation of the Cholesky stochastic volatility model is still an order of magnitude slower than that of the common stochastic volatility (when the natural conjugate prior is used). 

There are various useful extensions of the basic setup in \eqref{eq:VAR-SV}-\eqref{eq:hit}. First, instead of the independent autoregressive processes specified in \eqref{eq:hit}, $\bh_t$ can be modeled as a VAR(1) to allow for potential correlations among the volatility processes. In the extreme, \citet{CES20} consider a singular VAR for $\bh_t$ to capture the idea that only a few common shocks drive the evolution of the $n$ volatility processes. Second, the modified Cholesky factor $\bB_0$ in \eqref{eq:VAR-SV} can be made time-varying, as introduced in \citet{Primiceri05}. In fact, the multivariate stochastic volatility model of \citet{Primiceri05} is perhaps the most widely used specification in empirical macroeconomics, at least for small systems. 

Recent applications using large BVARs with the Cholesky stochastic volatility include \citet{BV18}, \citet{BGR18}, \citet{HF19}, \citet{CHP20}, \citet{KMMP20}, \citet{TZ20}, \citet{ZBZ20}, \citet{BKL22}, \citet{FHHP23} and \citet{GRR23}. 

\subsection{Factor Stochastic Volatility}

Another approach to model the law of motion of the error covariance matrix in \eqref{eq:VAR} is through the class of factor stochastic volatility models \citep{PS99b, AW00, CNS06}. A conventional factor model with homoskedastic factors and idiosyncratic errors is typically used for dimensionality reduction---to model linear relationships among a large number of variables in a very parsimonious way. For instance, an unrestricted $n\times n$ covariance matrix $\vSigma$ has $n(n+1)/2$ free elements; the number of parameters thus grows quadratically in $n$. To reduce the number of parameters to be estimated, a factor model in essence  specifies $\vSigma$ using simpler matrices $\bG, \bD$ and $\bL$ via $\vSigma = \bL\bG\bL' + \bD$, where $\bG$ and $\bD$ are $r\times r$ and $n\times n$ diagonal matrices, respectively, and $\bL$ is an $n\times r$ matrix, typically with $r\ll n$. Even for an unrestricted $\bL$, the number of parameters in a factor model is only $nr+n+r$; it grows linearly in $n$. Consequently, for high-dimensional settings with a large $n$ but a small $r$, a factor model can substantially reduce the number of parameters.

Factor stochastic volatility models extend the classic factor model by introducing time-varying volatility. An example is to specify the error covariance matrix $\vSigma_t$ as
\begin{equation} \label{eq:FSV}
	\vSigma_t = \bL\bG_t \bL' + \bD_t, 
\end{equation}
where $\bD_t = \text{diag}(\e^{h_{1,t}},\ldots, \e^{h_{n,t}})$ and $ \bG_t = \text{diag}(\e^{h_{n+1,t}},\ldots, \e^{h_{n+r,t}})$, and $h_{i,t}, i=1,\ldots, n+r,$ follows the AR(1) process in \eqref{eq:hit}. To facilitate estimation, data augmentation is typically used to represent the model in terms of latent factors. More specifically, consider the same VAR in \eqref{eq:VAR}, but now the innovation is constructed via 
\begin{equation}\label{eq:epsilon_decomposition}
	\vepsilon_t = \bL \mathbf{f}_t + \bu_t,
\end{equation}
where $\mathbf{f}_t = (f_{1,t},\ldots, f_{r,t})'$ is an $r\times 1$ vector of latent factors and
$\bL$ is the associated $n\times r$ factor loadings matrix. The errors $\bu_t$ and the latent factors $\mathbf{f}_t$ are assumed to be independent at all leads and lags and jointly Gaussian:
\begin{equation} \label{eq:ft}
	\begin{pmatrix}\bu_t \\  \mathbf{f}_t \end{pmatrix} \sim\distn{N}
	\left(\begin{pmatrix} \mathbf{0}\\ \mathbf{0} \end{pmatrix},
	\begin{pmatrix} \bD_t & \mathbf{0} \\ \mathbf{0} & \bG_t \end{pmatrix}\right),
\end{equation}
where $\bD_t = \text{diag}(\e^{h_{1,t}},\ldots, \e^{h_{n,t}})$ and $ \bG_t = \text{diag}(\e^{h_{n+1,t}},\ldots, \e^{h_{n+r,t}})$. Under the representation in \eqref{eq:epsilon_decomposition}-\eqref{eq:ft}, it is clear that the correlations among the elements of $\vepsilon_t$ are induced solely by the latent factors. 

For identification purposes, it is common to assume $\bL$ to be lower triangular with ones on the main diagonal. However, \citet{CEY22} show that in the presence of stochastic volatility and $r \leq (n-1)/2$, the factor loadings matrix $\bL$ is identified up to permutations and sign switches (with the restriction that $\mu_i, i=n+1,\ldots, n+r$, the unconditional means of the log-volatilities associated with the factors, are set to 0).\footnote{To handle the so-called label switching problem (the latent factors are permutations invariant), one can postprocess the posterior draws to sort them into the correct categories using, e.g., the approach in \citet{KS19}.}

In practice one also needs to determine the number of factors $r$. A common approach is to first fix the number of factors $r$ in the estimation, and then select $r$ using the marginal likelihood or some information criterion. An alternative approach is to include a large number of factors, and then use a shrinkage prior on the factor loadings so that they are increasingly shrunk to 0 as the column index increases, as considered in, e.g., \citet{BD11} and \citet{kastner19}. This latter approach circumvents the need to select the number of factors, but it makes the interpretation of the factors more difficult.

Compared to the Cholesky stochastic volatility, the factor stochastic volatility contains $r$ additional volatility processes and is apparently more flexible. However, as discussed earlier, and is clear in the decomposition in \eqref{eq:FSV}, the latter inherits the dimensionality reduction property of the classic factor model. As such, whether it fits macroeconomic time-series better than the two other stochastic volatility models is an empirical question. In terms of estimation, the factor stochastic volatility is more computationally intensive to estimate relative to the common stochastic volatility, but it can still be fitted reasonably quickly even when $n$ is large. In particular, given the latent factors, the VAR becomes $n$ unrelated regressions. Consequently, the $n$-equation system can be estimated equation by equation. 

While a variety of factor stochastic volatility models are commonly employed in financial applications, with recent examples in \citet{JMY19}, \citet{HV19}, \citet{kastner19}, \citet{LS20} and \citet{MMP20}, they are not yet widely used in the context of large BVARs. Notable exceptions are \citet{KH20} and \citet{HHKM22}.


\section{Order-Invariant and Outlier-Augmented Models} \label{s:OIOA}

In this section we introduce a variety of recently proposed stochastic volatility models with features that are especially relevant for large systems and for fitting post COVID-19 data.

\subsection{Order-Invariant Stochastic Volatility Models}

As outlined in Section~\ref{s:models}, the law of motion for $\vSigma_t$ is typically constructed by combining multiple volatility processes using some version of matrix decomposition or factorization. This construction process often leads to order dependence---estimates might depend on the order of variables arranged in the vector $\by_t$. For instance, the Cholesky stochastic volatility, which is based on a lower triangular parameterization, is not order invariant. In particular, as explained in \citet{CCM19}, since the model is constructed using a lower triangular impact matrix $\bB_0$ and the priors are independently elicited on $\bB_0$ and the volatility processes, the implied prior on $\vSigma_t$ is not order invariant. 

Even though this ordering issue is well-known and is explicitly discussed in \citet{CS05} and \citet{Primiceri05}, its empirical relevance has only been appreciated relatively recently. For example, \citet{ARRS22} use the model of \citet{Primiceri05} to produce point and density forecasts of 4 macroeconomic variables. They show that while the point forecasts are essentially the same across all the permutations of the 4 variables, the density forecasts can differ substantially across different variable orderings. Since the number of permutations grows exponentially as the number of variables increases, this ordering issue is expected to be more acute in large BVARs involving dozens of dependent variables. In fact, \citet{CDLS18} find evidence that estimates of reduced-form error variances based on the Cholesky stochastic volatility can drastically change across different variable orderings in large systems.\footnote{The structural analysis in both \citet{CS05} and \citet{Primiceri05} are based on a recursive identification scheme. As such, the order of the variables is part of the identification scheme and is therefore predetermined. However, many later studies \citep[e.g.,][]{benati2008, BP13} use the BVARs in \citet{CS05} and \citet{Primiceri05} as reduced-form models and implement other, non-recursive identification schemes. In those cases, the reduced-form models do not have a natural ordering. One such example is given in \citet{Bognanni18}, who demonstrates that responses to structural shocks identified with sign restrictions---using the reduced-form BVAR of \citet{Primiceri05}---can be sensitive to how the variables are ordered; see also \citet{Hartwig20} for another example.}

Given the sensitivity of the estimates and forecasts on different variable orderings, it is therefore desirable to employ order-invariant stochastic volatility models in empirical applications. Which of the stochastic volatility models discussed in Section~\ref{s:models} are order invariant? It is clear that the common stochastic volatility is order invariant, whereas, as discussed earlier, the Cholesky stochastic volatility is not. However, the latter can be extended to be order invariant by a slight modification: avoid the use of the lower triangular parameterization of $\bB_0$ in \eqref{eq:VAR-SV}. Based on this observation, \citet{CKY23} extend the Cholesky stochastic volatility model by allowing the impact matrix $\bB_0$ to be any dense, non-degenerate matrix. They prove that the model is order invariant, and based on the results in \citet{BB22}, the impact matrix $\bB_0$ is identified up to permutations and sign switches. Moreover, estimation can be done similarly as in the Cholesky stochastic volatility model, particularly that the VAR coefficients can be sampled equation by equation. \citet{CKY23} also consider a version of the model with a time-varying impact matrix $\bB_0$, thus extending the model of \citet{Primiceri05}. An alternative approach based on a different factorization of $\vSigma_t$ is considered in \citet{WK22}. More specifically, they construct $\vSigma_t$ using the eigendecomposition, where the $n$ log-eigenvalues evolve as random walks. They develop a fast posterior sampler to estimate the model and show that it is applicable to large datasets.

Finally, whether a factor stochastic volatility model is order invariant depends on the type of restrictions imposed on the factor loadings. As noted in \citet{kastner19}, if one imposes the usual identification restrictions that the factor loadings matrix is triangular, then the model is not order invariant. By contrast, \citet{CEY22} show that, under some mild conditions discussed in the previous section, a factor stochastic volatility model is order invariant and the factor loadings are identified up to permutations and sign switches.

There are other order-invariant modeling approaches, though they are typically designed for small systems. One popular approach is to construct multivariate stochastic volatility models using Wishart or inverse-Wishart processes; examples include \citet{PG2006} and \citet{AM09}. A few papers extend these multivariate stochastic volatility models to BVARs, including \citet{CDLS18}, \citet{SZ20} and \citet{ARRS22}. More recently, \citet{ARRS22} introduce a new order-invariant approach that builds on the new parameterization of correlation matrices proposed in \citet{AH21}. More specifically, they construct a time-varying correlation matrix based on random walk processes under the new parameterization. However, estimation of these stochastic volatility models is generally computationally intensive, as it often involves drawing from high-dimensional, non-standard distributions. Consequently, these stochastic volatility models do not scale well to large datasets.

Another order-invariant approach is based on the discounted Wishart process, which admits efficient filtering and smoothing algorithms for estimation. Stochastic volatility models constructed under this approach are considered in \citet{Uhlig97}, \citet{WH06}, \citet{PW10}, \citet{Bognanni18} and \citet{ARRS22}. However, the discounted Wishart process appears to be too tightly parameterized for typical macroeconomic data, and it tends to underperform in terms of both point and density forecasts relative to standard stochastic volatility models such as \citet{CS05} and \citet{Primiceri05}, as demonstrated in a forecasting exercise in \citet{ARRS22}.


%

\subsection{Outlier-Augmented Stochastic Volatility Models}

The COVID-19 pandemic has caused extreme movements in many macroeconomic and financial time-series. One consequence is that impulse response functions and forecasts from homoskedastic BVARs are heavily distorted by these extreme observations, as demonstrated in \citet{SS21} and \citet{LP22} using US data and \citet{BH23} using euro area data. This problem can be ameliorated by using BVARs with stochastic volatility, which downweight extreme observations. However, as pointed out by \citet{CCMM22}, extreme observations, by definition, should reflect transitory spikes, not permanent increases, in volatility. But in a typical stochastic volatility, changes in volatility are assumed to be highly persistent. It is therefore useful to augment standard stochastic volatility models to include an explicit component to model infrequent volatility spikes.

There is a general approach to directly model infrequent volatility spikes, and it can be incorporated into the many multivariate stochastic volatility models discussed earlier. The key idea is to replace the normal distribution in \eqref{eq:VAR} with a more robust distribution that puts more mass on extreme events. A common example is a continuous heavy-tailed distribution, such as the Student-$t$ distribution. As long as the robust distribution can be represented as a scale, finite or an infinite mixture of normals, estimation is straightforward thanks to data augmentation and the modular nature of MCMC methods. 

As an example, take the Cholesky stochastic volatility model with a generalized $t$ distribution considered in \citet{CR15}, \citet{CP16} and \citet{CMP17}, which is constructed from $n$ univariate Student-$t$ distributions with different degree of freedom parameters.\footnote{A similar heavy-tailed stochastic volatility model is introduced by \citet{CDG14} to construct a dynamic stochastic general equilibrium model that can accommodate extreme events during the Great Recession.} It can be represented as a conditionally Gaussian VAR specified in \eqref{eq:VAR} with $\vSigma_t$ constructed as
\begin{equation} \label{eq:SVt}
	\vSigma_t = \bB_0^{-1} \vLambda_t \bD_t (\bB_0^{-1})'.
\end{equation}
Compared to the standard Cholesky stochastic volatility in \eqref{eq:VAR-SV}, the new addition is the diagonal matrix of latent variables $\vLambda_t=\text{diag}(\lambda_{1,t},\ldots, \lambda_{n,t})$, where $\lambda_{i,t} \sim\distn{IG}(\nu_i/2, \nu_i/2)$ and $\distn{IG}(a, b)$ denotes the inverse-gamma distribution with mean $b/(a-1)$ when $a>1$.\footnote{For the special case with $\lambda_{1,t} = \cdots = \lambda_{n,t} = \lambda_t\sim\distn{IG}(\nu/2, \nu/2) $, the marginal distribution of the innovation $\vepsilon_t$ (unconditional on the latent variable $\lambda_t$) is a Student-$t$ distribution with degree of freedom parameter $\nu$ \citep*[see, e.g.,][]{geweke93, NS05}.} 

Another example is the outlier-augmented stochastic volatility model proposed by \citet{CCMM22}, which extends the Cholesky stochastic volatility by incorporating a discrete mixture representation first introduced in \citet{SW16} for handling outliers in unobserved components models. More specifically, it has the same conditionally Gaussian VAR representation with $\vSigma_t$ specified in \eqref{eq:SVt}, but the latent variable $\lambda_{i,t}$ is specified as $\lambda_{i,t} = o_{i,t}^2$, where $o_{i,t}$ follows a 2-part distribution with a point mass at 1 and a uniform distribution on the interval $(2,20)$. The first part of the distribution represents `regular' observations with scale normalized to 1, whereas the second part captures `outliers' that have 2-20 times larger standard deviations relative to regular observations.\footnote{In the original setup in \citet{SW16}, the range of the uniform distribution is from 2 to 10. As noted in \citet{CCMM22}, many macroeconomic variables exhibit extreme variability at the onset of the COVID-19 pandemic, and the upper range is accordingly increased to 20.}

Other distributional assumptions for $\lambda_{i,t}$ are possible and they would induce different marginal distributions on the innovations. Examples include the multivariate Laplace distribution \citep*{EKL06}, a finite mixture of scale mixtures of normals \citep{KM20} and an infinite mixture of normals \citep{Braun21}. These non-Gaussian distributions can also be incorporated in other stochastic volatility models, as is done in \citet{chan20} and \citet{Hartwig21}.

\section{Shrinkage Priors and Posterior Simulations} \label{s:estimation}

What makes BVARs Bayesian is the use of informative priors on the large number of VAR coefficients. Below we first outline a family of priors generally referred to as the Minnesota priors. We then discuss Bayesian estimation of BVARs with various types of stochastic volatility introduced earlier, particularly the posterior simulation of the log-volatilities and the VAR coefficients.

\subsection{Shrinkage  Priors on the VAR Coefficients}

We first discuss a family of priors on the VAR coefficients that can be traced back to the work in the 1980s by \citet*{DLS84} and \citet{litterman86}. Due to their affiliations at that time, this family of priors is commonly known as the Minnesota priors. In the original formulation for homoskedastic VARs, the error covariance matrix is fixed at some estimate, and a normal prior is elicited on the VAR coefficients $\valpha = \text{vec}(\bA')$. Later \citet{KK93, KK97} extend the original version to a joint prior on $\valpha$ and the time-invariant error covariance matrix. Since then many other more flexible variants have been developed, for both homoskedastic and heteroskedastic VARs, and they can be written as a conditionally normal prior given some latent variables or hyperparameters $\vxi$:
\begin{equation}\label{eq:prior}
	(\valpha\gvn \vxi) \sim \distn{N}(\valpha_0,\bV),
\end{equation}
where the prior covariance matrix $\bV$ may depend on $\vxi$. The prior mean $\valpha_0$ is typically set to zero for growth rates data, such as GDP growth. This reflects the prior belief that growth rates data are generally not very persistent, and the VAR coefficients are thus shrunk to 0. For levels data such as private investment or inflation indices, $\valpha_0$ is set to be zero except for the coefficients associated with the first own lag, which are set to be one. This reflects the prior belief that levels data are highly persistent---in particular, it expresses the preference for a random walk specification.

In early versions of the Minnesota priors there are no latent variables and the prior hyperparameters are fixed at some subjective values. The prior covariance matrix $\bV$ is typically assumed to be diagonal, and the diagonal elements are elicited to reflect a range of subjective beliefs, such as cross-variable shrinkage---i.e., the idea that coefficients on other variables' lags are on average smaller than coefficients on own lags. This is often implemented by introducing two hyperparameters, say, $\xi_1$ and $\xi_2$, where $\xi_1$ controls the overall prior variance for coefficients on own lags, whereas $\xi_2$ for coefficients on other lags. Then, $\xi_1$ and $\xi_2$ are carefully calibrated to incorporate the belief that $\xi_2$ is much smaller than~$\xi_1$.

Naturally, one set of fixed values for $\xi_1$ and $\xi_2$ is not expected to be optimal for all datasets with widely different time-series and sample periods. Indeed, \citet*{GLP15} demonstrate that substantial benefits can be obtained---e.g., better out-of-sample forecast performance using real data and more accurate impulse response functions using simulated data---simply by estimating these prior hyperparameters from the data instead of fixing them at some commonly-used subjective values. In a Bayesian model comparison exercise, \citet{chan23JE} finds strong evidence in favor of estimating these prior hyperparameters across BVARs with a variety of stochastic volatility specifications.\footnote{Earlier papers such as \citet{DS04} and \citet{CKM12} have adopted an empirical Bayes approach of selecting these prior hyperparameters by maximizing the marginal likelihood of the homoskedastic BVAR, which is available analytically. For time-varying parameter models, \citet{AMW20} show similar benefits in treating the prior hyperparameters as unknown parameters to be estimated.} 

There is a surge of interest in the statistics literature to develop hierarchical shrinkage priors, such as the Bayesian Lasso \citep{PC08}, the normal-gamma prior \citep{GB10}, the horseshoe prior \citep{CPS10horseshoe} and the Dirichlet-Laplace prior \citep{DLP15}. These shrinkage priors are originally designed for the setting of a linear regression with a large number of arbitrary predictors. More recently, they have been introduced in BVAR settings. Applications include \citet{HF19}, \citet{FY19}, \citet{KH20}, \citet{KP19} and \citet{GKP23}. In contrast to Minnesota priors that tend to shrink all VAR coefficients, these adaptive hierarchical priors have the desirable theoretical property of only shrinking `small' coefficients strongly to zero, while leaving `large' coefficients mostly intact. 

In practice, however, these new hierarchical priors do not seem to forecast better than some sophisticated Minnesota priors, as demonstrated in \citet{CHP20}. One reason for this surprising result could be because the new hierarchical priors typically treat all VAR coefficients identically, and do not take into account the many plausible prior beliefs, such as cross-variable shrinkage and that variables of higher lags are less important, incorporated into the Minnesota priors. There are a few papers that aim to adapt the new adaptive hierarchical priors so that they are more suitable for BVARs. For instance, \citet{HF19} develop a normal-gamma prior in which additional lag-specific shrinkage parameters are introduced with the goal of shrinking higher-order lags more strongly to zero. \citet{KP19} consider a class of hierarchical shrinkage priors that incorporate some features of the Minnesota priors. \citet{chan21} develops a framework that explicitly nests both the Minnesota priors and the new adaptive hierarchical priors. These Minnesota-type adaptive hierarchical priors can be written as conditionally Gaussian priors with additional latent variables $\vxi$ given in \eqref{eq:prior}. For example, under the Minnesota-type normal-gamma prior $\vxi$ becomes a vector of gamma random variables; for the Minnesota-type horseshoe prior, elements of $\vxi$ have a half-Cauchy distribution. \citet{chan21} shows that these Minnesota-type adaptive hierarchical priors forecast better than both the conventional Minnesota priors and the standard adaptive hierarchical priors.

The literature on shrinkage priors on VAR coefficients is vast and expanding. Here we have focused on the family of Minnesota priors. More discussions on other shrinkage priors and approaches can be found in Chapter 2 of this volume by Hauzenberger, Huber and Koop and excellent reviews by \citet{KK10} and \citet{karlsson13}. 


\subsection{Bayesian Estimation}

Estimating BVARs with stochastic volatility tends to be computationally intensive, especially for large systems. But a lot of progress has been made in the last decade, and it is now feasible to fit large BVARs with a variety of stochastic volatility specifications using MCMC methods. Below we discuss the posterior simulation of the two most computationally intensive parts: sampling the log-volatilities and the VAR coefficients.

\subsubsection{Posterior Simulation of the Log-Volatilities}

Conditional on the data and other model parameters, the VAR in \eqref{eq:VAR} defines the observation equation for the log-volatilities. But since the log-volatilities enter the model via the time-varying covariance matrix, the state space representation is nonlinear and high-dimensional, which makes estimation very challenging in general. However, many multivariate stochastic volatility models, such as common, Cholesky and factor stochastic volatility, are constructed in such a way that they can be transformed into a linear (though non-Gaussian) state space model, for which the auxiliary mixture sampler of \citet{KSC98} can be applied to sample the log-volatilities efficiently. 

More specifically, after transforming the original nonlinear Gaussian representation into a linear non-Gaussian state space model, \citet{KSC98} observe that the non-Gaussian errors in the new observation equation can be well approximated using a 7-component mixture of normals. Then, given the mixture indicators, the approximate model becomes a linear Gaussian state space model for which sampling the latent states can be done using standard algorithms, such as the Kalman-filter based smoothers of \citet{CK94} and \citet{DK02} or the precision-based samplers of \citet{CJ09} and \citet{MMP11}. Estimation details of the common, Cholesky and factor stochastic volatility models can be found in \citet{chan23JE}. Other stochastic volatility models with additional features can also be estimated using the auxiliary mixture sampler with minor modifications; see, for example, \citet{CCMM22} and \citet{CKY23}. 

One important point to emphasize is that, as pointed out in \citet{DP15}, when the auxiliary mixture sampler is used as part of a larger posterior simulator, the latter becomes a collapsed Gibbs sampler, where the augmented mixture indicators and all other model parameters should be sampled jointly. This joint sampling can be done by first drawing the model parameters given the log-volatilities but marginal of the mixture indicators, followed by sampling the mixture indicators given the log-volatilities and other model parameters. In contrast to a standard Gibbs sampler, here the sampling order is important.\footnote{It is also feasible to apply a standard Gibbs sampler by conditioning on the mixture indicators in all the Gibbs steps. This approach is typically more cumbersome as, e.g., the VAR in \eqref{eq:VAR} is no longer Gaussian given the mixture indicators. See \citet{SW16} for an example of this alternative approach.} In practice, the main implication is that the log-volatilities should be sampled immediately after the mixture indicators.\footnote{For stochastic volatility models with multiple time-series of log-volatilities, where each series has an associated vector of mixture indicators, it is often more convenient to sample each pair of mixture indicators and log-volatilities separately. This approach is valid if the time-series of log-volatilities are conditionally independent given the model parameters---an example is the Cholesky stochastic volatility model specified in \eqref{eq:VAR-SV}-\eqref{eq:hit}. If they are not conditionally independent, then all the mixture indicators should be sampled first before the log-volatilities.}

For other stochastic volatility models for which the auxiliary mixture sampler cannot be applied, there are two alternatives. While they are both very general and can be applied to a wide range of stochastic volatility models, they are more computationally demanding. The first is to use sequential Monte Carlo (SMC) or particle filters. There is a large and rapidly expanding literature on sequential Monte Carlo methods; \citet{DDG01} and \citet{Creal12} provide excellent reviews on this large family of algorithms. The most widely used member is perhaps the bootstrap particle filter, where the state equation is used to generate particle values from time $t$ to values at time~$t+1$. While the bootstrap particle filter is easy to implement, its empirical performance is often poor. The main reason is that the generated particle values from the state equation are unlikely to be consistent with the data in typical settings. There are a number of more sophisticated filters that use information from the observation equation to construct better proposal distributions. A recent example is the tempered particle filter proposed in \citet{HS19} that adaptively constructs the proposal distribution through a sequence of tempering steps.

The second general approach is to implement a Metropolis-Hastings step to sample the log-volatilities. This can be done by using either a random-walk Metropolis-Hastings algorithm as in \citet{AM09}, or an independence-chain Metropolis-Hastings algorithm with a tailored proposal distribution. The first option is straightforward to implement, but is often very inefficient in terms of highly autocorrelated posterior draws, especially in high-dimensional settings. The second option requires a good proposal distribution that well approximates the conditional distribution of the log-volatilities. As such, it demands more analysis and is therefore more difficult to implement. However, there has been a lot of progress in developing good proposal distributions. For example, \citet{McCausland12} introduces the HESSIAN method based on a fifth-order Taylor approximation that works remarkably well for a range of univariate stochastic volatility models. \citet{CE18} and \citet{CHKP23} instead consider quadratic (hence, Gaussian) approximations of the target distributions in the context of BVARs with stochastic volatility, which are easier to obtain and work well for their applications.\footnote{This approach of using Gaussian approximations is first introduced to estimate univariate stochastic volatility models in the seminal papers by \citet{DK97} and \citet{SP97}. They formulate the approximating algorithm via a linear Gaussian state space model, which implies a Gaussian density for the states. \citet{CE18} and \citet{CHKP23} improve upon this approach by directly computing the Gaussian approximating density using Newton-Raphson method based on fast band matrix routines.} In addition, sequential Monte Carlo methods can also be used to construct good proposal distributions; see, e.g., \citet{ADH10}.
 

\subsubsection{Posterior Simulation of the VAR Coefficients}

Given the conditionally Gaussian VAR in \eqref{eq:VAR} and the conditionally Gaussian prior on the VAR coefficients $\valpha$ in \eqref{eq:prior}, the conditional posterior distribution of $\valpha$ is also Gaussian by standard linear regression results. Hence, in principle sampling $\valpha$ is straightforward. In practice, however, when the number of endogenous variables in the VAR is large, conventional ways of drawing all VAR coefficients jointly---e.g., by first obtaining the Cholesky factor of the covariance matrix or the precision matrix---become excessively computationally intensive. In particular, for an $n$-variable VAR with $p$ lags, the number of VAR coefficients (ignoring intercepts) is $n^2p$. Since the covariance matrix of the joint distribution is of dimension $n^2p\times n^2p$, obtaining its Cholesky factor has computational complexity of the order $\mathcal{O}(n^6p^3)$.\footnote{Sampling the VAR coefficients using the Cholesky factor of the precision matrix is faster, as it avoids explicitly inverting the precision matrix to obtain the covariance matrix. But since the precision matrix is dense in this case, its computational complexity is of the same order of $\mathcal{O}(n^6p^3)$.} 

To tackle this computational problem, \citet*{CCM19} consider an alternative sampling scheme of drawing the VAR coefficients equation by equation that can drastically reduce the computational complexity by two orders of magnitude to $\mathcal{O}(n^4p^3)$.\footnote{As noted in \citet{Bognanni22}, the algorithm in \citet{CCM19} can only be viewed as an approximation. \citet{CCCM22} provide an exact algorithm to draw the VAR coefficients equation by equation that has the same order of computational complexity.} The key reason for the speed-up is as follows. Since the number of VAR coefficients in each equation is $np$, drawing these coefficients conditional on coefficients in other equations can be done in $\mathcal{O}(n^3p^3)$ operations. Iterating this step over $n$ equations, the total computational complexity is therefore  $\mathcal{O}(n^4p^3)$. This equation-by-equation estimation algorithm is formulated for the Cholesky stochastic volatility model, but it can be used for other types of stochastic volatility or homoskedastic VARs with minor modifications.\footnote{For some special cases, other faster algorithms are available. For example, under the natural conjugate prior, the common stochastic volatility model and its variants can be estimated using algorithms with complexity $\mathcal{O}(n^3p^3)$; see \citet{CCM16} and \citet{chan20}.} \citet{CCCM22} report a 10-50 times speed-up of the equation-by-equation approach for $n$ up to 40 compared to sampling all coefficients jointly.

Another venue to improve sampling speed is to find faster ways to simulate from high-dimensional Gaussian distributions. For a linear regression with $r$ observations and $m$ covariates, the conventional, Cholesky-factor based sampling approach to simulate the $m$-vector of coefficients requires $\mathcal{O}(m^3)$ operations. \citet{BCM16} propose an alternative way to sample the coefficients based on the Woodbury formula, which has computational complexity of $\mathcal{O}(r^2m)$. As such, this new algorithm is expected to be faster in cases where there are far more covariates than observations. For our BVAR setting with $r=Tn$ and $m= np$ (sampling the coefficients in one equation), the new algorithm has computational complexity of $\mathcal{O}(T^2 n^3p)$ compared to the conventional sampling approach of $\mathcal{O}(n^3p^3)$. Both have the same order of complexity in $n$, but they have different complexity in other dimensions. For instance, when $T$ is small and $p$ is large, the algorithm of \citet{BCM16} is expected to perform better.

\subsubsection{Alternative Bayesian Approaches}

So far we have focused on various MCMC methods to estimate the reduced-form VAR given in \eqref{eq:VAR}. There is a range of additional strategies to speed up computations so that one can fit larger BVARs with more flexible features.

One approach is to reparameterize the standard reduced-form VAR in the structural form with a recursive system \citep[see, e.g.,][]{chan21, GKP23}. The structural-form parameter estimates can then be transformed to obtain the corresponding reduced-form estimates. The main  advantage of this approach is that the $n$-variable VAR can be directly written as a system of $n$ unrelated univariate regressions, which can be estimated separately. This reparameterization can offer up to 10 times speed up compared to estimation of the reduced form, depending on the size of the system. Moreover, further speed-up can be achieved via parallelization. The key drawback of this approach is that to apply standard conjugacy results, one needs to use normal priors on the structural-form VAR coefficients, which often imply order-dependent priors on the reduced-form parameters. One way to ameliorate this ordering issue is to first elicit prior beliefs on the reduced-form parameters, and then work out the implied prior beliefs on the structural-form parameters; see, e.g., \citet{chan22} for an example.

There are other Bayesian estimation approaches that offer substantial reduction in computation time compared to conventional MCMC methods. For example, \citet{BZ20} introduce a sequential Monte Carlo (SMC) algorithm for estimating BVARs with stochastic volatility that leverages parallelization. In addition, the algorithm is especially suitable for online learning tasks, such as out-of-sample forecasting, where new data can be incorporated to update the posterior distribution without completely reestimating the model.

While MCMC and SMC are exact methods in the sense that they can approximate the posterior distribution arbitrarily well by increasing the simulation size, there are approximate methods available in settings where neither MCMC nor SMC are computationally feasible. One prominent example is the fast growing collection of variational Bayesian methods, which are deterministic algorithms for approximating the posterior distribution using a more tractable distribution. More specifically, given a family of tractable densities, the variational Bayesian approach locates the optimal density within this family by minimizing the Kullback-Leibler divergence of the approximating density to the posterior density. Key references on variational Bayesian methods can be found in \citet{Jordanetal99}, \citet{Bishop06} and \citet{OW10}.

\citet{HW18} appear to be the first to apply the variational Bayesian approach to the estimation of a 7-variable homoskedastic VAR. \citet{CY22} and \citet{GKP23} develop variational Bayesian methods for estimating large BVARs with stochastic volatility. In particular, \citet{CY22} use a 96-variable BVAR to measure global bank network connectedness, whereas \citet{GKP23} conduct an out-of-sample forecasting exercise using a large data set of 100 variables. While both \citet{CY22} and \citet{GKP23} parameterize their BVARs in the structural form, \citet{BBB23} develop a variational Bayes algorithm to estimate a reduced-form BVAR with stochastic volatility. Hierarchical shrinkage priors on the reduced-form VAR coefficients seem to perform better than those elicited on the structural form, but estimation under the reduced-form parameterization is slightly more costly.



\section{Time-Varying Parameter and Nonlinear VARs} \label{s:nonlinear}

Since the pioneering works of \citet{CS01,CS05} and \citet{Primiceri05}, BVARs with time-varying VAR coefficients have been widely used to model time-series with structural changes. In these time-varying parameter (TVP) models, it is crucial to also allow for time-varying volatility---as highlighted in \citet{Sims01}, a failure to account for heteroskedasticity in a TVP model with constant error variances might give an appearance of time variation in the VAR coefficients. In fact, \citet{SZ06} compare various regime-switching BVARs and find that the best model is the one that allows for time variation in error variances only. \citet{CE18} report similar findings in BVARs with TVP and stochastic volatility.

Of course, TVP models are but one nonlinear approach to model the conditional mean in BVARs. For applications that study the impact of macroeconomic or financial uncertainty on key macroeconomic variables, BVARs with stochastic volatility in mean are an especially convenient framework. For example, \citet{MZ13} develop a TVP-VAR with stochastic volatility where the log-volatilities and their lags enter the conditional mean as regressors. Using this new model, they then study the impact of monetary shocks volatility. An alternative approach based on a large BVAR with a multiplicative factor stochastic volatility model is proposed in \citet{CCM18}. In their model the log-volatilities are driven by common factors representing macroeconomic
and financial uncertainty. The two common factors and their lags then enter the conditional mean in the BVAR as regressors. Estimation of these stochastic volatility in mean models is typically more computationally intensive, as the auxiliary mixture sampler does not apply. In view of this difficulty, \citet{CHKP23} develop an efficient precision-based sampler for posterior and predictive inference in large stochastic volatility in mean BVARs. They report substantial computational and statistical efficiency gains over a standard particle filter.

There are also a few nonparametric approaches designed for BVARs. For example, \citet{HHMP21} develop a nonparametric BVAR using a Gaussian process prior to model the relationship of each endogenous variable on its own lags and lags of all other variables. The model also features a multivariate stochastic volatility specification similar to the Cholesky stochastic volatility. An alternative nonparametric approach based on Bayesian additive regression trees is considered in \citet{HHKM22} and \citet{ClarkEtal22}. Specifically, a TVP-VAR with stochastic volatility is first represented in the structural form, and in each equation both the time-varying conditional mean and the conditional variance are modeled using Bayesian additive regression trees. Both find forecast accuracy gains over linear BVARs. For more detailed discussions on Bayesian nonparametric methods, see Chapter 5 of this volume by Marcellino and Pfarrhofer.


\section{The Road Ahead} \label{s:conclusions}

Over the last decade substantial progress has been made in developing a wide range of stochastic volatility models and efficient estimation approaches for large datasets. There remains, however, a lot of ongoing work in building more flexible time-varying BVARs. One direction tackles the problem of modeling time variation in large systems, since standard TVP models of \citet{CS05} and \citet{Primiceri05} cannot be directly applied to those settings due to computational and overparameterization concerns. In addition, allowing all VAR coefficients to be time-varying appears to be unnecessary. For example, \citet{chan23} finds evidence that in a large BVAR, the coefficients in some, but not all, equations are time-varying, and forcing all coefficients to be time-varying worsens forecast performance. Papers such as \citet{KK13, KK18} propose fast estimation methods to approximate the posterior distributions of large TVP models. \citet{HKO19} develop an algorithm that first shrinks the time-varying coefficients, followed by setting the small values to zero. The nonparametric approach based on Bayesian additive regression trees developed in \citet{HHKM22} can also accommodate time-varying coefficients. 


Another research direction involves developing more general VAR models able to accommodate the augmentation of typical macroeconomic datasets with additional types of data. A promising recent approach is the functional VAR developed in \citet{CCS21} that jointly models standard macroeconomic time-series and individual-level cross-sectional data. This framework is especially suitable for studying, for example, the distributional effects of various structural macroeconomic shocks.

\newpage

\singlespace

\bibliographystyle{econometrica}
\bibliography{VAR}

\end{document}